\documentclass[aps,prl,twocolumn,nofootinbib,floatfix,preprintnumbers]{revtex4}
\usepackage{subfigure}
\usepackage{graphicx}
\usepackage{dcolumn}
\usepackage{epsfig}
\usepackage{amsmath}
\usepackage{amsfonts}
\usepackage{amssymb}
\usepackage{color}
\usepackage{hyperref}
\usepackage[normalem]{ulem}

\usepackage[T1]{fontenc}
\usepackage{ae,aecompl}
\def\be{\begin{equation}}
\def\ee{\end{equation}}
\def\ba{\begin{eqnarray}}
\def\ea{\end{eqnarray}}

\newcommand{\bx}{{\bf x}}

\newcommand{\bq}{{\bf q}}

\def\bx{{\bf x}}         
\def\bq{{\bf q}}         
\def\everything{V}       

\begin{document}

\title{Optimal Transport Reconstruction of Baryon Acoustic Oscillations}

\author{Farnik Nikakhtar}
\email{farnik@sas.upenn.edu}
\affiliation{Department of Physics and Astronomy, University of Pennsylvania, Philadelphia, PA 19104 -- USA}
\affiliation{Department of Physics, Yale University, New Haven, CT 06511, USA}

\author{Ravi K.~Sheth}
\affiliation{Center for Particle Cosmology, University of Pennsylvania, Philadelphia, PA 19104 -- USA}
\affiliation{The Abdus Salam International Center for Theoretical Physics, Strada Costiera 11, Trieste 34151 -- Italy}

\author{Bruno L\'evy}
\affiliation{Centre Inria de Paris, 2 Rue Simone Iff, 75012 Paris, France}
 
\author{Roya Mohayaee}
\affiliation{Sorbonne Universit\'e, CNRS, Institut d'Astrophysique de Paris, 98bis Bld Arago, 75014 Paris -- France}

\affiliation{Rudolf Peierls Centre for Theoretical Physics, University of Oxford, Parks Road, Oxford OX1 3PU -- United Kingdom}


\begin{abstract}
 A weighted, semi-discrete, fast optimal transport (OT) algorithm for reconstructing the Lagrangian positions of proto-halos from their evolved Eulerian positions is presented.  The algorithm makes use of a mass estimate of the biased tracers and of the distribution of the remaining mass (the `dust'), but is robust to errors in the mass estimates.  Tests with state-of-art cosmological simulations show that if the dust is assumed to have a uniform spatial distribution, then the shape of the OT-reconstructed pair correlation function of the tracers is very close to linear theory, enabling sub-percent precision in the BAO distance scale that depends weakly, if at all, on a cosmological model.  With a more sophisticated model for the dust, OT returns an estimate of the displacement field which yields superb reconstruction of the proto-halo positions, and hence of the shape {\em and} amplitude of the initial pair correlation function of the tracers.  This enables direct and independent determinations of the bias factor $b$ and the smearing scale $\Sigma$, potentially providing new methods for breaking the degeneracy between $b$ and $\sigma_8$.  
\end{abstract}


\maketitle

The Baryon Acoustic Oscillations (BAO) are frozen sound waves from the pre-decoupling era that leave a peak in the 2-point correlation function (2PCF) of the linear matter distribution \cite{1970ApJ...162..815P} on scales of order 140~Mpc. This `BAO peak' provides an important distance scale for measuring the expansion history of the Universe \cite{2005ApJ...633..560E}.  However, gravitational evolution shifts and smears this peak in the late-time baryon plus dark matter distribution \cite{Bharadwaj1996, rpt, esw2007}.  This potentially biases the inferred distance scale, so most BAO analyses aim to sharpen the peak and remove its shift by undoing the effects of the gravity \cite{2007ApJ...664..675E, recSDSS}.  This `reconstruction' is complicated by the fact that we only observe biased tracers of the full density field.  Therefore, most density field reconstruction methods make assumptions about the background cosmology and the growth of perturbations in it, as well as the nature of the bias between the observed tracers and the dark matter \cite{Sarpa:2018ucb, recIterate, recHE}. In what follows, we describe and test a method in which such assumptions enter much more weakly.

\begin{figure*}
 \centering
 \includegraphics[width=1\textwidth]{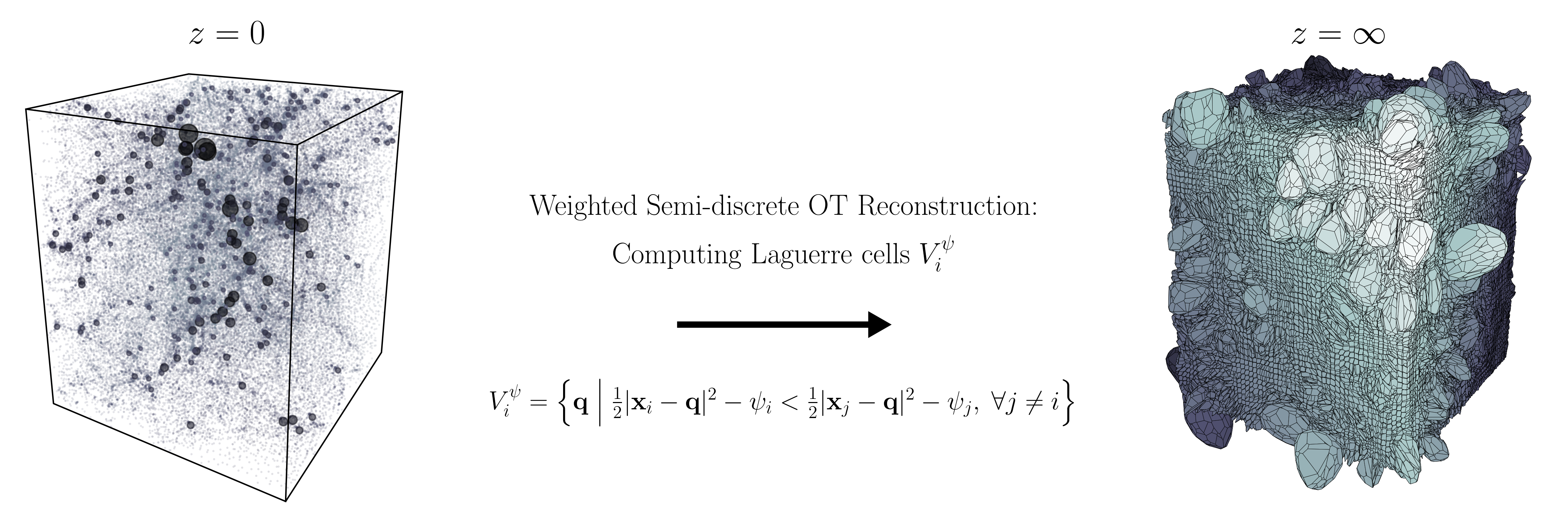}
 \caption{\label{fig:laguerre-Schema} Schema showing how our fast, weighted, semi-discrete algorithm reconstructs Laguerre cells from the present distribution of biased tracers (halos) and dark matter field particles (dust). The volume of each Laguerre cell represents the mass of the object (halo or dust) to which it corresponds. The initial power spectrum/2PCF is obtained directly from the distribution of the barycenters of the Laguerre cells; there is no need to make any additional cosmological model-dependent assumptions.}
\end{figure*}

Optimal transport (hereafter OT) is a powerful mathematical framework which has recently found applications in diverse branches of science \cite{BenamouBrenier, leonard, MondinoSuhr, PeyreCuturi}.  In the present context, OT is a deterministic algorithm that recovers the initial Lagrangian positions $\bq$ of a given final Eulerian distribution of particles $\bx$ by solving the Monge-Amp\`ere-Kantorovich (MAK) problem \cite{EURNature, eur03}.  In semi-discrete OT \cite{royaMAK, royaMAK_PRL}, the final particle distribution is considered to have evolved from a smooth, uniform, continuous initial field (rather than a discrete cartesian grid). In this setting, instead of a single point, a patch of Lagrangian space, {\it a Laguerre cell}, is assigned to each evolved object. The map that assigns the Laguerre cells to final Eulerian positions is the unique solution to a semi-discrete optimization problem as well \cite{BrenierPFMR91}; it is specified by that set of $\psi_i$ values which maximizes the `Kantorovich dual' \cite{OTON, Santambrogio, DBLP:journals/cg/LevyS18}:  
\be
    K(\psi) = \sum_i \int_{\everything^\psi_i} \left[\frac{1}{2}|\bx_i - \bq|^2 - \psi_i\right]\; d^3q + \sum_i  v_i\psi_i,
    \label{eq:Kpsi}
\ee
subject to $V^\psi_i \neq \emptyset$ for all $i$. Equation~\ref{eq:Kpsi} assumes that the Lagrangian density field is uniform, so $v_i$ is the volume of the $i$th Laguerre cell, imposed as a constraint (the $v_i$ sum up to the total volume).  The $\psi_i$ coefficients correspond to the Lagrange multiplier associated with the constraint (if all $\psi_i$ are equal, the Laguerre diagram reduces to the Voronoi diagram \cite{Voronoi94}).  Whereas previous semi-discrete OT reconstructions used the same $v_i$ for all cells \cite{royaMAK}, the {\em weighted} semi-discrete OT we use here allows $v_i\ne v_j$ (see Figure~\ref{fig:laguerre-Schema}).  
 
The objective function $K(\psi)$ is concave -- guaranteeing the existence and uniqueness of $\psi$ -- and smooth ($C^2$).  Therefore, the exhaustive combinatorial computation of the discrete-discrete Monge problem can be replaced with an efficient and convergent Newton method
\cite{DBLP:journals/corr/KitagawaMT16, DBLP:conf/compgeom/AurenhammerHA92, journals/M2AN/LevyNAL15}.
Hence, our semi-discrete approach scales as $\mathcal{O}\left( N_{\rm LC}\log N_{\rm LC}\right)$ \cite{royaMAK}, where $N_{\rm LC}$ is the number of Laguerre cells: a {\em significant} improvement over previous algorithms.

For the dark matter field, comprised of equal mass particles, the assumption of uniform initial conditions implies that all the $v_i$ in equation~\ref{eq:Kpsi} are equal. The associated OT algorithm accurately recovers the set of displacements $\bx-\bq$ which maps the $z=0$ position $\bx$ of each dark matter particle to the barycenter of its associated Laguerre cell $\bq$ \cite{royaMAK}. 
Treating biased tracers, and halos in particular, is more complicated because:
\begin{itemize}
    \item[i.] Halos typically span a range of masses.  
    \item[ii.] Whereas the initial dark matter field was uniform, the initial proto-halo distribution was not.  Moreover, the Lagrangian-space clustering of the proto-halo patches depends on their mass \cite{mw1996, st1999}.  
\end{itemize}
Accommodating (i) is straightforward:  The assumption of uniform initial conditions means that one simply sets $v_i\propto m_i$ in equation~(\ref{eq:Kpsi}), where $m_i=n_i\,m_p$ is the mass of each object ($m_p$ is the particle mass). 
However, in principle, (ii) seriously complicates the OT approach.  To address this, we include a model for the Eulerian `dust' -- the mass that is not associated with the biased tracers -- since the initial distribution of (mass-weighted) tracers plus dust should be uniform.  Typically (e.g. for BAO surveys), the biased tracers account for only $\sim 20\%$ of the total mass density.  Fortunately, our OT algorithm is fast, so the additional computational load required to reconstruct dust as well as the biased tracers is offset by the simplicity of the initial condition. 

We present results for two dust models:  
\begin{itemize}
  \item[U.]  Dust is uniformly distributed; if the total mass in dust is $M_d$, we use $N_d$ particles each of mass $m_d = M_d/N_d$ to model it.   
\end{itemize}
\begin{itemize}
  \item[W.]  The dust is arranged in a cosmic `web', represented by a random subset of the dark matter particles in a simulation which were not assigned to the biased tracers. 
\end{itemize}
Implementing U in real data is straightforward; W will be more complicated.  

\begin{figure*}
 \centering
  \includegraphics[width=1\textwidth]{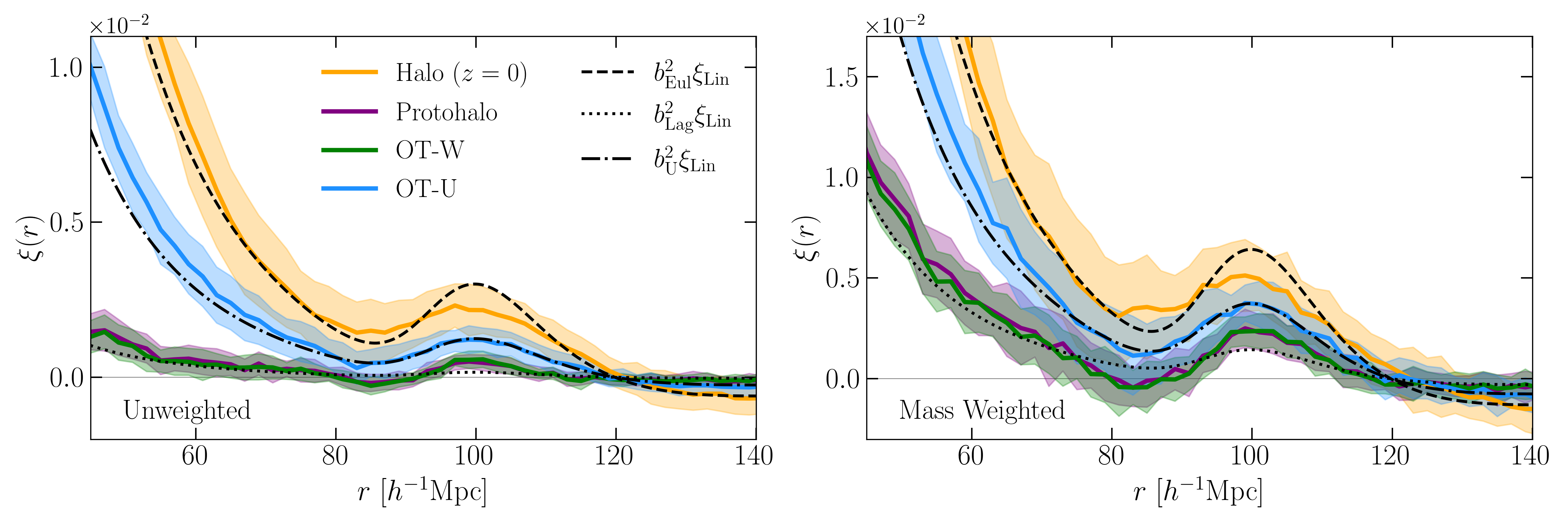}
 \caption{\label{fig:xi} Number- (left) and mass-weighted (right) 2PCFs measured in non-overlapping bins of width $2h^{-1}$Mpc. Thick orange curve shows $\xi_{\rm Eul}$ of the $z=0$ halo centers averaged over 20 simulations, and yellow band shows the rms scatter around this mean.  Thick purple curve and error band shows $\xi_{\rm Lag}$ of the corresponding Lagrangian (i.e. $z=\infty$) protohalo centers.  Thick blue curve and error band shows $\xi_{\rm\text{OT-U}}$ measured in our OT reconstructed field when the Eulerian dust is assumed to be uniformly distributed.  Thick green curve and error band -- which is very similar to $\xi_{\rm Lag}$ -- shows $\xi_{\rm\text{OT-W}}$ measured in our OT reconstructed field when the dust (correctly) traces the cosmic web.  Smooth dashed, dotted and dot-dashed curves show $b^2\xi_{\rm Lin}$, $(b-1)^2\xi_{\rm Lin}$ and $b_{\rm U}^2\xi_{\rm Lin}$, where $\xi_{\rm Lin}$ is the 2PCF of the dark matter in linear theory.  
 In both panels, the difference in shape between $\xi_{\rm Eul}$ and $b^2\xi_{\rm Lin}$ is why reconstruction is necessary. Our reconstructed $\xi_{\rm\text{OT-U}}$ reproduces the shape of $\xi_{\rm Lin}$ around the BAO feature very well.  The difference in shape between $\xi_{\rm Lag}$ and $(b-1)^2\xi_{\rm Lin}$, which $\xi_{\rm\text{OT-W}}$ reproduces exquisitely, is due to scale dependent Lagrangian bias.  
 } 
\end{figure*}

Although both dust models make an assumption about the background density, the results to follow are robust to $\sim 10$\% variations in this density.  Since the background density is already known to much higher precision, other than when converting angles and redshifts to distances, our method requires no other assumption about the background cosmological model.  

In particular, since the OT reconstructed field is uniform, all previous 
OT work measured spatial statistics using $\bq + D(z)\,(\bx-\bq)$, for some $D(z) \ll 1$, rather than $\bq$ itself.  Since this resembles the linear Lagrangian `Zel'dovich' approximation, the analysis appeared to be cosmology dependent, even though the OT step did not assume a cosmology. 
In contrast, the Lagrangian distribution of biased tracers is non-trivial, so the BAO feature can be obtained directly from the OT-reconstructed positions $\bq$.\footnote{We show elsewhere that this can be done even for the dark matter if one estimates the Lagrangian 2PCF by weighting each Laguerre cell by the divergence of $\mathbf{x-q}$, or by plotting the Laplacian of the correlation function of the displacement field.}  

We demonstrate our results using halos identified in 20 realizations of the HADES simulations of \cite{hades}. 
Each simulation follows the gravitational evolution of $512^3$ identical particles, each of mass $m_p=6.566\times 10^{11}h^{-1}M_\odot$, in a periodic cube of side $L=1h^{-1}$Gpc in which the background cosmology is flat, $\Omega_\Lambda = 1-\Omega_m$, with $(\Omega_m,\Omega_b)=(0.3175,0.049)$, and Hubble constant today $H_0=100h$~km~s$^{-1}$Mpc$^{-1}$, with $h=0.6711$.  The initial fluctuation field is Gaussian so it is completely specified by its power spectrum, $P_{\rm Lin}(k)$, which is taken from CLASS, with shape $n_s=0.9624$ and amplitude set by $\sigma_8=0.833$.  
Halos were identified in the $z=0$ output of each box using a friends-of-friends algorithm with linking length $b=0.2$.  We only use halos with more than 20 particles.
About 75\% percent of the particles in the box are not bound up in such halos:  these make up the `dust'.  

Our OT algorithm takes as input the list of halo masses (results are robust to factor-of-2 or 3 errors) and positions and a model for the spatial distribution of the dust.  For (W), we select dust particles randomly with probability $p$; our results (which used $p=1$) are not sensitive to its value.  
For (U), we simply randomize the positions of these dust particles.  

The list of $n_i$ particles which makes up a halo of mass $m_i=n_i\,m_p$ defines a protohalo patch in the initial conditions.  We refer to the center of mass of this halo at the time it was identified ($z=0$) as its Eulerian position $\bx_i$, and that of the protohalo as the Lagrangian position $\bq_i$ ($z=\infty$; strictly speaking the initial conditions are at $z_{\rm IC}=99$, but the $z_{\rm IC}$ to $\infty$ displacement is negligible).  

Figure~\ref{fig:xi} compares 2PCFs averaged over 20 simulation boxes.  
In each panel, the solid orange curve shows the average halo correlation function $\xi_{\rm Eul}(r)$ (i.e. using the Eulerian positions $|\mathbf{x}_i-\mathbf{x}_j|$), and the yellow bands around it show the standard deviation.  The purple curve shows $\xi_{\rm Lag}$ of the Lagrangian protohalos (which is built from the $|\mathbf{q}_i-\mathbf{q}_j|$ pairs). The blue and green curves show $\xi_{\rm\text{OT-U}}$ and $\xi_{\rm\text{OT-W}}$, which are built from the OT-reconstructed Laguerre cell barycenters with the two dust models (U and W).

To show that our results do not depend on the halo sample, in the right hand panel we weighted each halo by its mass when computing the pair counts (this weight is the same for the corresponding protohalos and Laguerre cells). Additionally, mass-weighting halos (above some threshold) gives a good first approximation for galaxy, as opposed to halo, pair counts \cite{halomodel}. Comparison with the left hand panel shows that mass-weighting increases the amplitudes of all the 2PCFs, as expected \cite{st1999}.\footnote{The mass-weighted correlation function of all the particles -- i.e. including the `dust' -- is zero, confirming that, once all the mass is included, the OT algorithm has indeed converged to a uniform density initial condition.} Except for this, the qualitative trends in the two panels are the same:  $\xi_{\rm Eul} > \xi_{\rm\text{OT-U}} > \xi_{\rm Lag}$ and $\xi_{\rm\text{OT-W}}\approx \xi_{\rm Lag}$.

We are most interested in the shapes of these $\xi$, since differences in amplitude alone will not bias cosmological constraints.  Therefore, the smooth curves in each panel show $b^2\xi_{\rm Lin}$, for three choices of the `linear', `scale-independent' bias factor $b$.  The dashed (upper most) curve has $b_{\rm Eul} = 1.4$ in the left hand panel and 1.9 in the right.
This shows clearly that, compared to the shape of $\xi_{\rm Lin}$, the BAO feature is smeared out in $\xi_{\rm Eul}$ (orange).  This is well understood \cite{Bharadwaj1996,rpt,esw2007}, and is why reconstruction is necessary.  

The dot-dashed curves have $b_{\rm U} = 0.84$ and 1.45 in the two panels, and show that $\xi_{\rm\text{OT-U}}$ (blue) is {\em extremely} close in shape to $\xi_{\rm Lin}$ (on BAO scales).  In fact, OT-U has recovered the linear theory shape better than all published reconstruction algorithms to date.  

\begin{figure}
 \centering
 \includegraphics[width=0.45\textwidth]{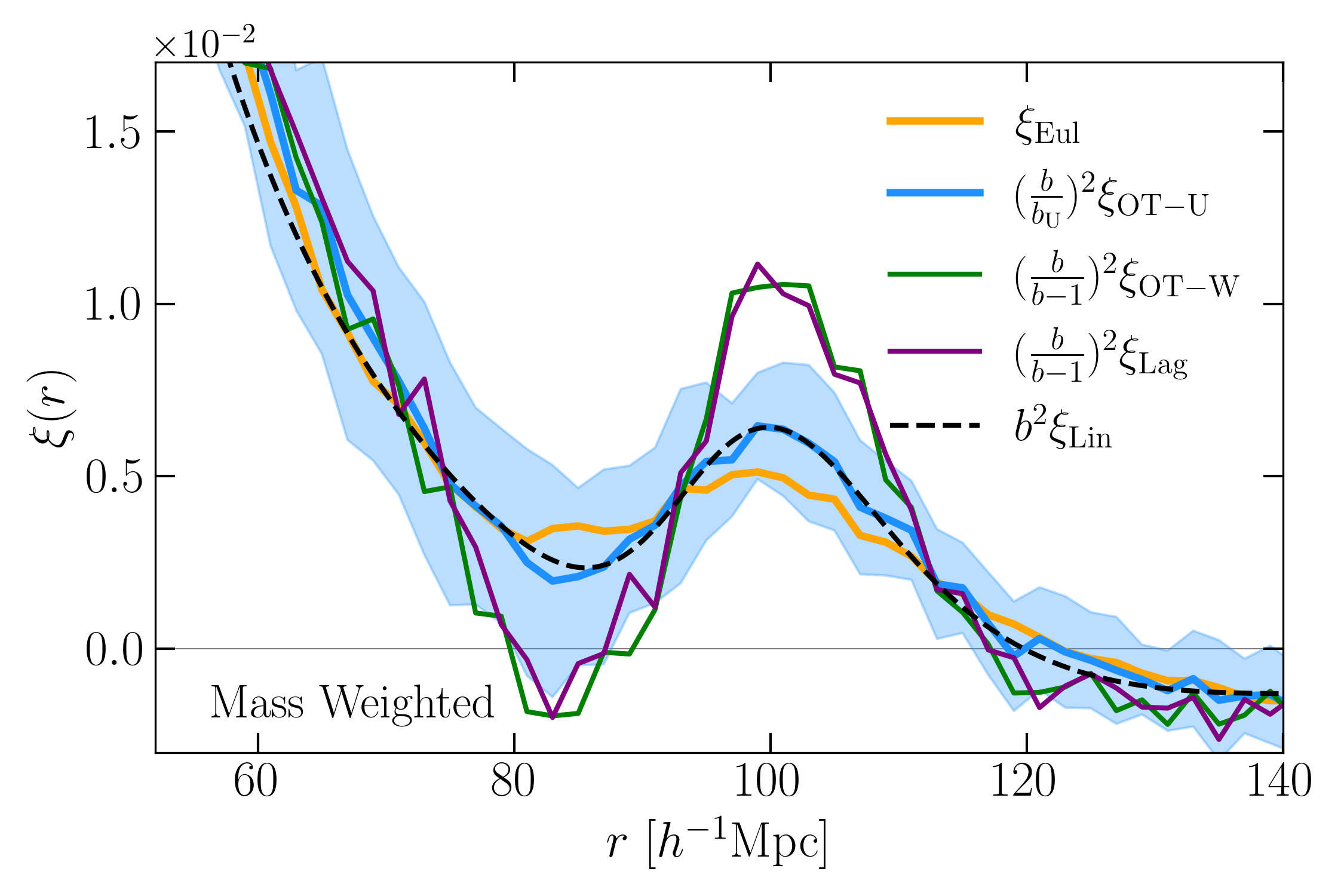}
 \caption{\label{fig:scaled} Same as right hand panel of Figure~\ref{fig:xi}, but now yellow, blue, green and purple curves show the mean of $\xi_{\rm Eul}$, $(b/b_{\rm U})^2\,\xi_{\rm\text{OT-U}}$, $[b/(b-1)]^2\,\xi_{\rm\text{OT-W}}$ and $[b/(b-1)]^2\,\xi_{\rm Lag}$, and we only show error bands around the blue (OT-U) curve.  All curves intersect at approximately the same `Linear Point' scale.
 }
\end{figure}

Finally, the dotted curves have $b_{\rm Lag} = b_{\rm Eul} - 1$ because, for tracers of a fixed number density, such as Eulerian halos and their Lagrangian protohalos, if 
\begin{equation}
  \xi_{\rm Eul}\approx b_{\rm Eul}^2\,\xi_{\rm Lin} ,\quad 
  {\rm then}
  \quad \xi_{\rm Lag}\approx (b_{\rm Eul}-1)^2\,\xi_{\rm Lin}
  \label{eq:bevolve}
\end{equation}
\cite{nd1994,fry1996,mw1996,st1999}. 
These curves show that, while the amplitude of $\xi_{\rm Lag}$ (purple) scales as expected, its shape is typically {\em more} peaked than $\xi_{\rm Lin}$.  The reason for this `scale-dependent bias' is well understood \cite{bbks86, bkPeaks, baoPeaks}, but this shape difference is ignored -- or simply not reproduced -- by essentially all other reconstruction schemes \cite[e.g.][]{recSDSS,recIterate,recHE,eFAM2019}.  In contrast, $\xi_{\rm\text{OT-W}}$ (green) reproduces $\xi_{\rm Lag}$ (purple) on these scales exquisitely, whatever the value of $b_{\rm Eul}$.  
Figure~\ref{fig:scaled} highlights this by showing $\xi_{\rm Eul}$, $b^2\xi_{\rm Lin}$, $[b/(b-1)]^2\xi_{\rm Lag}$,  $[b/(b-1)]^2\xi_{\rm\text{OT-W}}$ and $(b/b_{\rm U})^2\xi_{\rm\text{OT-U}}$.

The agreement between $\xi_{\rm\text{OT-W}}$ and $\xi_{\rm Lag}$ is a significant success.  However, since $\xi_{\rm\text{OT-W}}$ (like $\xi_{\rm Lag}$) differs in shape from $\xi_{\rm Lin}$, we now address how to estimate the BAO scale. (Note that this is not an issue for $\xi_{\rm\text{OT-U}}$.) 

The traditional approach fits a fiducial $\xi_{\rm Lin}$ template shape to the reconstructed 2PCF, but as $\xi_{\rm\text{OT-W}}$ and $\xi_{\rm Lin}$ have different shapes, this might lead to a bias. Instead, `peaks theory' \cite{bbks86, bkPeaks, baoPeaks} provides a much better fiducial template; we report on its use elsewhere.  The second, motivated by \cite{recSDSS, baoPeaks}, corrects the shape of $\xi_{\rm\text{OT-W}}$ so that it becomes more like $\xi_{\rm Lin}$ by using the OT-W displacements to distort a uniform grid.  
The third, described below, does {\em not} require prior knowledge of $\xi_{\rm Lin}$.

Figure~3 shows that $\xi_{\rm Eul}$, $b^2\xi_{\rm Lin}$, $[b/(b-1)]^2\xi_{\rm Lag}$,  $[b/(b-1)]^2\xi_{\rm\text{OT-W}}$ and $(b/b_{\rm U})^2\xi_{\rm\text{OT-U}}$ intersect at a scale that is approximately half-way in between the peak and dip scales.  Evidently, 
 $r_{\rm LP}\equiv (r_{\rm peak} + r_{\rm dip})/2$,
is relatively immune to both evolution {\em and} scale dependent bias \cite{rLP2016}, so we will use it to quantify the gain from using $\xi_{\rm\text{OT-W}}$ rather than $\xi_{\rm Eul}$.
 
To estimate $r_{\rm LP}$ we fit polynomials, or more carefully chosen basis functions, to $\xi$ over the range $r=[60,120]h^{-1}$Mpc, and then differentiate the fit \cite{lpMocks, laguerre, hermite}.  The peak and LP scales in $\xi_{\rm Eul}$, $(r_{\rm peak},r_{\rm LP})/h^{-1}$Mpc = (98.5, 91.1), are shifted by 1\% from their values in $\xi_{\rm Lin}$, (99.8, 92.7); this is why reconstruction was necessary.  In contrast, they are (100.7, 92.6) in both $\xi_{\rm Lag}$ and $\xi_{\rm\text{OT-W}}$, and (99.8, 92.6) in $\xi_{\rm\text{OT-U}}$. Evidently, (i) $r_{\rm LP}$ is indeed insensitive to scale-dependent bias, and (ii) OT reconstruction provides a nearly unbiased constraint on the cosmological distance scale.

Distance scale estimates use the shape of $\xi_{\rm OT}$ but not its amplitude.  In most analyses, this amplitude is proportional to the product of $b\sigma_8$, where $\sigma_8$ is the amplitude of fluctuations in the dark matter field.  The fidelity and robustness of $\xi_{\rm\text{OT-W}}$ potentially enable a number of novel analyses for breaking the $b\sigma_8$ degeneracy.  

Analysis I is more standard:  Ratios of the 2PCF and 3PCFs yield $b$ independent of $\sigma_8$ \cite{Bernardeau:2001qr}.  In the Eulerian field these ratios involve terms coming from evolution as well as from bias.  However, since our OT-W reconstructions are so faithful to the Lagrangian field, measuring these ratios in the OT-W field is both feasible and easy to interpret (no evolution, only bias).  Moreover, combining them with similar measurements in the evolved field should provide useful constraints on the contribution from evolution.

Analysis II:  Because $\xi_{\rm Eul}/\xi_{\rm\text{OT-W}} \approx (1 - 1/b)^{-2}$ on $\sim 50$-$70h^{-1}$Mpc scales (Figure~\ref{fig:xi}), one could combine measurements over a range of $r$ by fitting for that multiplicative factor $A$ which brings $\xi_{\rm Eul}$ and $\xi_{\rm\text{OT-W}}$ into agreement over this range of scales.  Since $A=(1-1/b)^2$ this furnishes a direct estimate of $b$ with {\em no} assumptions about the shape or amplitude of $\xi_{\rm Lin}$. 

\begin{figure}
 \centering
 \includegraphics[width=0.8\hsize]{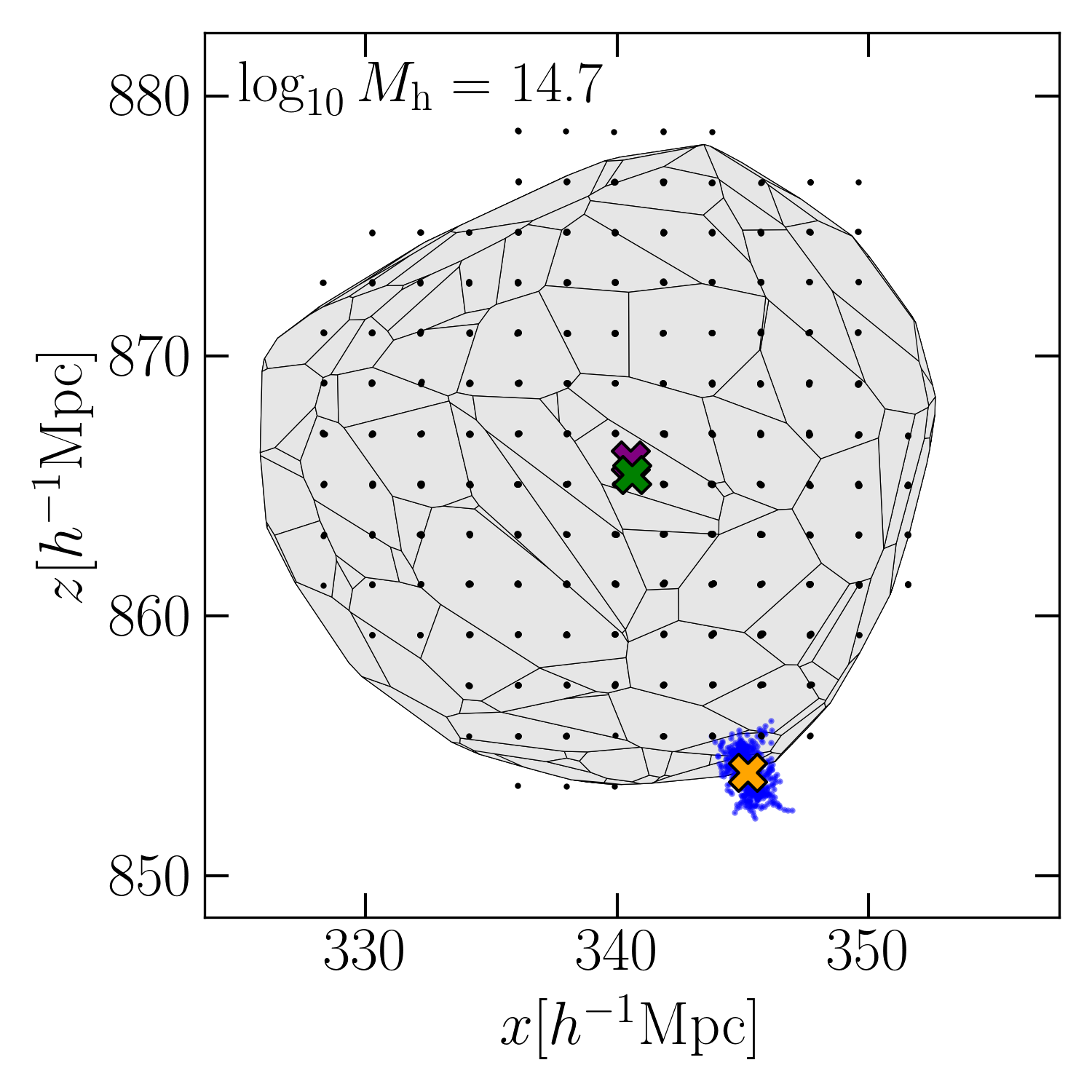}
 \caption{\label{fig:protohalos} Eulerian distribution of particles that make up a halo (upper left corner shows the mass in $[h^{-1}M_{\odot}]$) at $z=0$ (blue dots), and Lagrangian (i.e. $z=\infty$) distribution of the same particles (grid-like black dots); yellow and purple crosses show their centers-of-mass are displaced.  
 Green cross shows the barycenter of the OT-W reconstructed Laguerre cell (grey).
 }
\end{figure}

Analysis III:  Figure~\ref{fig:protohalos} shows that OT-W also reconstructs the shape and position of each protohalo patch quite well.  Reconstructing the shapes is non-trivial \cite{shapes2013, lbp2014}, but here we focus on the positions. Figure~\ref{fig:protohalos} shows that the barycenter of the Laguerre cell $\bq_{\rm\text{OT-W}}$ is extremely close to the actual protohalo $\bq$, even though both are offset from the Eulerian $\bx$.  It is because $|\bx-\bq|\ne 0$ that the BAO feature in $\xi_{\rm Eul}$ is smeared out compared to $\xi_{\rm Lag}$, and because $\bx-\bq \approx \bx-\bq_{\rm\text{OT-W}}$ for all halos, $\xi_{\rm\text{OT-W}}\approx \xi_{\rm Lag}$.  Therefore, $\Sigma_{\rm\text{OT-W}}$, the rms of $|\bx-\bq_{\rm\text{OT-W}}|$, must be very similar to $\Sigma_{\rm disp}$, the rms of $|\bx-\bq|$.  If we ignore redshift space distortions, $\Sigma_{\rm disp}^2\approx \int dk\,P_{\rm Lin}(k)/2\pi^2$ is independent of $b$ \cite{laguerre, hermite}.  Following Ref.~\cite{hermite}, let $\Sigma_{\rm obs}^2$ denote the result of performing this integral with the observed $P_{\rm Eul}(k)$ in place of $P_{\rm Lin}(k)$.  Since $P_{\rm Eul}\approx b^2\,P_{\rm Lin}$, the ratio $\Sigma_{\rm obs}/\Sigma_{\rm\text{OT-W}}\approx b$.\footnote{N.B. $(2/3)\Sigma^2_{\rm\text{OT-W}}$ can be used directly in the Laguerre reconstruction algorithm of Refs.~\cite{laguerre, hermite} to provide a consistency check of the BAO distance scale.}  In practice, errors on the input halo masses will impact $\Sigma_{\rm\text{OT-U}}$ and $\Sigma_{\rm\text{OT-W}}$, and hence the estimated $b$.  Quantifying this, 
extending OT to work with redshift-space distorted positions (following, e.g., \cite{royaMAK, mt2005, 2massMAK}), and modeling the dust (e.g., \cite{cai2011, elucid2014, voronoi2020}), before applying it to data, are all work in progress.

The authors thank Enzo Branchini for discussions. F. N. acknowledges the support of the Balzan Foundation via the Balzan Fellowship in final stages of this work.

\bibliography{PRLrev.bib}

\begin{thebibliography}{48}
\expandafter\ifx\csname natexlab\endcsname\relax\def\natexlab#1{#1}\fi
\expandafter\ifx\csname bibnamefont\endcsname\relax
  \def\bibnamefont#1{#1}\fi
\expandafter\ifx\csname bibfnamefont\endcsname\relax
  \def\bibfnamefont#1{#1}\fi
\expandafter\ifx\csname citenamefont\endcsname\relax
  \def\citenamefont#1{#1}\fi
\expandafter\ifx\csname url\endcsname\relax
  \def\url#1{\texttt{#1}}\fi
\expandafter\ifx\csname urlprefix\endcsname\relax\def\urlprefix{URL }\fi
\providecommand{\bibinfo}[2]{#2}
\providecommand{\eprint}[2][]{\url{#2}}

\bibitem[{\citenamefont{{Peebles} and {Yu}}(1970)}]{1970ApJ...162..815P}
\bibinfo{author}{\bibfnamefont{P.~J.~E.} \bibnamefont{{Peebles}}}
  \bibnamefont{and} \bibinfo{author}{\bibfnamefont{J.~T.} \bibnamefont{{Yu}}},
  \bibinfo{journal}{\apj} \textbf{\bibinfo{volume}{162}}, \bibinfo{pages}{815}
  (\bibinfo{year}{1970}).

\bibitem[{\citenamefont{{Eisenstein} et~al.}(2005)\citenamefont{{Eisenstein},
  {Zehavi}, {Hogg}, {Scoccimarro}, {Blanton}, {Nichol}, {Scranton}, {Seo},
  {Tegmark}, {Zheng} et~al.}}]{2005ApJ...633..560E}
\bibinfo{author}{\bibfnamefont{D.~J.} \bibnamefont{{Eisenstein}}},
  \bibinfo{author}{\bibfnamefont{I.}~\bibnamefont{{Zehavi}}},
  \bibinfo{author}{\bibfnamefont{D.~W.} \bibnamefont{{Hogg}}},
  \bibinfo{author}{\bibfnamefont{R.}~\bibnamefont{{Scoccimarro}}},
  \bibinfo{author}{\bibfnamefont{M.~R.} \bibnamefont{{Blanton}}},
  \bibinfo{author}{\bibfnamefont{R.~C.} \bibnamefont{{Nichol}}},
  \bibinfo{author}{\bibfnamefont{R.}~\bibnamefont{{Scranton}}},
  \bibinfo{author}{\bibfnamefont{H.-J.} \bibnamefont{{Seo}}},
  \bibinfo{author}{\bibfnamefont{M.}~\bibnamefont{{Tegmark}}},
  \bibinfo{author}{\bibfnamefont{Z.}~\bibnamefont{{Zheng}}},
  \bibnamefont{et~al.}, \bibinfo{journal}{\apj} \textbf{\bibinfo{volume}{633}},
  \bibinfo{pages}{560} (\bibinfo{year}{2005}), \eprint{astro-ph/0501171}.

\bibitem[{\citenamefont{{Bharadwaj}}(1996)}]{Bharadwaj1996}
\bibinfo{author}{\bibfnamefont{S.}~\bibnamefont{{Bharadwaj}}},
  \bibinfo{journal}{\apj} \textbf{\bibinfo{volume}{472}}, \bibinfo{pages}{1}
  (\bibinfo{year}{1996}), \eprint{astro-ph/9606121}.

\bibitem[{\citenamefont{{Crocce} and {Scoccimarro}}(2008)}]{rpt}
\bibinfo{author}{\bibfnamefont{M.}~\bibnamefont{{Crocce}}} \bibnamefont{and}
  \bibinfo{author}{\bibfnamefont{R.}~\bibnamefont{{Scoccimarro}}},
  \bibinfo{journal}{\prd} \textbf{\bibinfo{volume}{77}}, \bibinfo{eid}{023533}
  (\bibinfo{year}{2008}), \eprint{0704.2783}.

\bibitem[{\citenamefont{{Eisenstein}
  et~al.}(2007{\natexlab{a}})\citenamefont{{Eisenstein}, {Seo}, and
  {White}}}]{esw2007}
\bibinfo{author}{\bibfnamefont{D.~J.} \bibnamefont{{Eisenstein}}},
  \bibinfo{author}{\bibfnamefont{H.-J.} \bibnamefont{{Seo}}}, \bibnamefont{and}
  \bibinfo{author}{\bibfnamefont{M.}~\bibnamefont{{White}}},
  \bibinfo{journal}{\apj} \textbf{\bibinfo{volume}{664}}, \bibinfo{pages}{660}
  (\bibinfo{year}{2007}{\natexlab{a}}), \eprint{astro-ph/0604361}.

\bibitem[{\citenamefont{{Eisenstein}
  et~al.}(2007{\natexlab{b}})\citenamefont{{Eisenstein}, {Seo}, {Sirko}, and
  {Spergel}}}]{2007ApJ...664..675E}
\bibinfo{author}{\bibfnamefont{D.~J.} \bibnamefont{{Eisenstein}}},
  \bibinfo{author}{\bibfnamefont{H.-J.} \bibnamefont{{Seo}}},
  \bibinfo{author}{\bibfnamefont{E.}~\bibnamefont{{Sirko}}}, \bibnamefont{and}
  \bibinfo{author}{\bibfnamefont{D.~N.} \bibnamefont{{Spergel}}},
  \bibinfo{journal}{\apj} \textbf{\bibinfo{volume}{664}}, \bibinfo{pages}{675}
  (\bibinfo{year}{2007}{\natexlab{b}}), \eprint{astro-ph/0604362}.

\bibitem[{\citenamefont{{Padmanabhan} et~al.}(2012)\citenamefont{{Padmanabhan},
  {Xu}, {Eisenstein}, {Scalzo}, {Cuesta}, {Mehta}, and {Kazin}}}]{recSDSS}
\bibinfo{author}{\bibfnamefont{N.}~\bibnamefont{{Padmanabhan}}},
  \bibinfo{author}{\bibfnamefont{X.}~\bibnamefont{{Xu}}},
  \bibinfo{author}{\bibfnamefont{D.~J.} \bibnamefont{{Eisenstein}}},
  \bibinfo{author}{\bibfnamefont{R.}~\bibnamefont{{Scalzo}}},
  \bibinfo{author}{\bibfnamefont{A.~J.} \bibnamefont{{Cuesta}}},
  \bibinfo{author}{\bibfnamefont{K.~T.} \bibnamefont{{Mehta}}},
  \bibnamefont{and} \bibinfo{author}{\bibfnamefont{E.}~\bibnamefont{{Kazin}}},
  \bibinfo{journal}{\mnras} \textbf{\bibinfo{volume}{427}},
  \bibinfo{pages}{2132} (\bibinfo{year}{2012}), \eprint{1202.0090}.

\bibitem[{\citenamefont{Sarpa et~al.}(2019)\citenamefont{Sarpa, Schimd,
  Branchini, and Matarrese}}]{Sarpa:2018ucb}
\bibinfo{author}{\bibfnamefont{E.}~\bibnamefont{Sarpa}},
  \bibinfo{author}{\bibfnamefont{C.}~\bibnamefont{Schimd}},
  \bibinfo{author}{\bibfnamefont{E.}~\bibnamefont{Branchini}},
  \bibnamefont{and}
  \bibinfo{author}{\bibfnamefont{S.}~\bibnamefont{Matarrese}},
  \bibinfo{journal}{\mnras} \textbf{\bibinfo{volume}{484}},
  \bibinfo{pages}{3818} (\bibinfo{year}{2019}), \eprint{1809.10738}.

\bibitem[{\citenamefont{{Schmittfull} et~al.}(2017)\citenamefont{{Schmittfull},
  {Baldauf}, and {Zaldarriaga}}}]{recIterate}
\bibinfo{author}{\bibfnamefont{M.}~\bibnamefont{{Schmittfull}}},
  \bibinfo{author}{\bibfnamefont{T.}~\bibnamefont{{Baldauf}}},
  \bibnamefont{and}
  \bibinfo{author}{\bibfnamefont{M.}~\bibnamefont{{Zaldarriaga}}},
  \bibinfo{journal}{\prd} \textbf{\bibinfo{volume}{96}}, \bibinfo{eid}{023505}
  (\bibinfo{year}{2017}), \eprint{1704.06634}.

\bibitem[{\citenamefont{{Hada} and {Eisenstein}}(2018)}]{recHE}
\bibinfo{author}{\bibfnamefont{R.}~\bibnamefont{{Hada}}} \bibnamefont{and}
  \bibinfo{author}{\bibfnamefont{D.~J.} \bibnamefont{{Eisenstein}}},
  \bibinfo{journal}{\mnras} \textbf{\bibinfo{volume}{478}},
  \bibinfo{pages}{1866} (\bibinfo{year}{2018}), \eprint{1804.04738}.

\bibitem[{\citenamefont{Benamou and Brenier}(2000)}]{BenamouBrenier}
\bibinfo{author}{\bibfnamefont{J.}~\bibnamefont{Benamou}} \bibnamefont{and}
  \bibinfo{author}{\bibfnamefont{Y.}~\bibnamefont{Brenier}},
  \bibinfo{journal}{Numerische Mathematik} \textbf{\bibinfo{volume}{84}},
  \bibinfo{pages}{375} (\bibinfo{year}{2000}),
  \urlprefix\url{https://doi.org/10.1007/s002110050002}.

\bibitem[{\citenamefont{L\'eonard}(2014)}]{leonard}
\bibinfo{author}{\bibfnamefont{C.}~\bibnamefont{L\'eonard}},
  \bibinfo{journal}{Discrete \& Continuous Dynamical Systems}
  \textbf{\bibinfo{volume}{34}}, \bibinfo{pages}{1533} (\bibinfo{year}{2014}).

\bibitem[{\citenamefont{{Mondino} and {Suhr}}(2018)}]{MondinoSuhr}
\bibinfo{author}{\bibfnamefont{A.}~\bibnamefont{{Mondino}}} \bibnamefont{and}
  \bibinfo{author}{\bibfnamefont{S.}~\bibnamefont{{Suhr}}},
  \bibinfo{journal}{arXiv e-prints} \bibinfo{eid}{arXiv:1810.13309}
  (\bibinfo{year}{2018}), \eprint{1810.13309}.

\bibitem[{\citenamefont{{Peyr{\'e}} and {Cuturi}}(2018)}]{PeyreCuturi}
\bibinfo{author}{\bibfnamefont{G.}~\bibnamefont{{Peyr{\'e}}}} \bibnamefont{and}
  \bibinfo{author}{\bibfnamefont{M.}~\bibnamefont{{Cuturi}}},
  \bibinfo{journal}{arXiv e-prints} \bibinfo{eid}{arXiv:1803.00567}
  (\bibinfo{year}{2018}), \eprint{1803.00567}.

\bibitem[{\citenamefont{Frisch et~al.}(2002)\citenamefont{Frisch, Matarrese,
  Mohayaee, and Sobolevskii}}]{EURNature}
\bibinfo{author}{\bibfnamefont{U.}~\bibnamefont{Frisch}},
  \bibinfo{author}{\bibfnamefont{S.}~\bibnamefont{Matarrese}},
  \bibinfo{author}{\bibfnamefont{R.}~\bibnamefont{Mohayaee}}, \bibnamefont{and}
  \bibinfo{author}{\bibfnamefont{A.}~\bibnamefont{Sobolevskii}},
  \bibinfo{journal}{Nature} \textbf{\bibinfo{volume}{417}}
  (\bibinfo{year}{2002}).

\bibitem[{\citenamefont{Brenier et~al.}(2003)\citenamefont{Brenier, Frisch,
  H{\'e}non, Loeper, Matarrese, Mohayaee, and Sobolevskii}}]{eur03}
\bibinfo{author}{\bibfnamefont{Y.}~\bibnamefont{Brenier}},
  \bibinfo{author}{\bibfnamefont{U.}~\bibnamefont{Frisch}},
  \bibinfo{author}{\bibfnamefont{M.}~\bibnamefont{H{\'e}non}},
  \bibinfo{author}{\bibfnamefont{G.}~\bibnamefont{Loeper}},
  \bibinfo{author}{\bibfnamefont{S.}~\bibnamefont{Matarrese}},
  \bibinfo{author}{\bibfnamefont{R.}~\bibnamefont{Mohayaee}}, \bibnamefont{and}
  \bibinfo{author}{\bibfnamefont{A.}~\bibnamefont{Sobolevskii}},
  \bibinfo{journal}{Mon. Not. R. Astron. Soc.} \textbf{\bibinfo{volume}{346}}
  (\bibinfo{year}{2003}).

\bibitem[{\citenamefont{{Levy} et~al.}(2021)\citenamefont{{Levy}, {Mohayaee},
  and {von Hausegger}}}]{royaMAK}
\bibinfo{author}{\bibfnamefont{B.}~\bibnamefont{{Levy}}},
  \bibinfo{author}{\bibfnamefont{R.}~\bibnamefont{{Mohayaee}}},
  \bibnamefont{and} \bibinfo{author}{\bibfnamefont{S.}~\bibnamefont{{von
  Hausegger}}}, \bibinfo{journal}{\mnras} \textbf{\bibinfo{volume}{506}},
  \bibinfo{pages}{1165} (\bibinfo{year}{2021}), \eprint{2012.09074}.

\bibitem[{\citenamefont{{von Hausegger} et~al.}(2022)\citenamefont{{von
  Hausegger}, {L{\'e}vy}, and {Mohayaee}}}]{royaMAK_PRL}
\bibinfo{author}{\bibfnamefont{S.}~\bibnamefont{{von Hausegger}}},
  \bibinfo{author}{\bibfnamefont{B.}~\bibnamefont{{L{\'e}vy}}},
  \bibnamefont{and}
  \bibinfo{author}{\bibfnamefont{R.}~\bibnamefont{{Mohayaee}}},
  \bibinfo{journal}{\prl} \textbf{\bibinfo{volume}{128}}, \bibinfo{eid}{201302}
  (\bibinfo{year}{2022}), \eprint{2110.08868}.

\bibitem[{\citenamefont{Brenier}(1991)}]{BrenierPFMR91}
\bibinfo{author}{\bibfnamefont{Y.}~\bibnamefont{Brenier}},
  \bibinfo{journal}{Communications on Pure and Applied Mathematics}
  \textbf{\bibinfo{volume}{44}}, \bibinfo{pages}{375} (\bibinfo{year}{1991}).

\bibitem[{\citenamefont{Villani}(2009)}]{OTON}
\bibinfo{author}{\bibfnamefont{C.}~\bibnamefont{Villani}},
  \emph{\bibinfo{title}{Optimal transport : old and new}}, Grundlehren der
  mathematischen Wissenschaften (\bibinfo{publisher}{Springer},
  \bibinfo{address}{Berlin}, \bibinfo{year}{2009}), ISBN
  \bibinfo{isbn}{978-3-540-71049-3},
  \urlprefix\url{http://opac.inria.fr/record=b1129524}.

\bibitem[{\citenamefont{Santambrogio}(2015)}]{Santambrogio}
\bibinfo{author}{\bibfnamefont{F.}~\bibnamefont{Santambrogio}},
  \emph{\bibinfo{title}{Optimal transport for applied mathematicians}},
  vol.~\bibinfo{volume}{87} of \emph{\bibinfo{series}{Progress in Nonlinear
  Differential Equations and their Applications}}
  (\bibinfo{publisher}{Birkh\"auser/Springer, Cham}, \bibinfo{year}{2015}),
  ISBN \bibinfo{isbn}{978-3-319-20827-5; 978-3-319-20828-2},
  \bibinfo{note}{calculus of variations, PDEs, and modeling}.

\bibitem[{\citenamefont{L{\'{e}}vy and
  Schwindt}(2018)}]{DBLP:journals/cg/LevyS18}
\bibinfo{author}{\bibfnamefont{B.}~\bibnamefont{L{\'{e}}vy}} \bibnamefont{and}
  \bibinfo{author}{\bibfnamefont{E.~L.} \bibnamefont{Schwindt}},
  \bibinfo{journal}{Comput. Graph.} \textbf{\bibinfo{volume}{72}},
  \bibinfo{pages}{135} (\bibinfo{year}{2018}),
  \urlprefix\url{https://doi.org/10.1016/j.cag.2018.01.009}.

\bibitem[{\citenamefont{{van de Weygaert}}(1994)}]{Voronoi94}
\bibinfo{author}{\bibfnamefont{R.}~\bibnamefont{{van de Weygaert}}},
  \bibinfo{journal}{\aap} \textbf{\bibinfo{volume}{283}}, \bibinfo{pages}{361}
  (\bibinfo{year}{1994}).

\bibitem[{\citenamefont{Kitagawa et~al.}(2016)\citenamefont{Kitagawa,
  M{\'{e}}rigot, and Thibert}}]{DBLP:journals/corr/KitagawaMT16}
\bibinfo{author}{\bibfnamefont{J.}~\bibnamefont{Kitagawa}},
  \bibinfo{author}{\bibfnamefont{Q.}~\bibnamefont{M{\'{e}}rigot}},
  \bibnamefont{and} \bibinfo{author}{\bibfnamefont{B.}~\bibnamefont{Thibert}},
  \bibinfo{journal}{CoRR}  (\bibinfo{year}{2016}), \eprint{1603.05579},
  \urlprefix\url{http://arxiv.org/abs/1603.05579}.

\bibitem[{\citenamefont{Aurenhammer et~al.}(1992)\citenamefont{Aurenhammer,
  Hoffmann, and Aronov}}]{DBLP:conf/compgeom/AurenhammerHA92}
\bibinfo{author}{\bibfnamefont{F.}~\bibnamefont{Aurenhammer}},
  \bibinfo{author}{\bibfnamefont{F.}~\bibnamefont{Hoffmann}}, \bibnamefont{and}
  \bibinfo{author}{\bibfnamefont{B.}~\bibnamefont{Aronov}}, in
  \emph{\bibinfo{booktitle}{Symposium on Computational Geometry}}
  (\bibinfo{year}{1992}), pp. \bibinfo{pages}{350--357}.

\bibitem[{\citenamefont{L{\'{e}}vy}(2015)}]{journals/M2AN/LevyNAL15}
\bibinfo{author}{\bibfnamefont{B.}~\bibnamefont{L{\'{e}}vy}},
  \bibinfo{journal}{ESAIM M2AN (Mathematical Modeling and Analysis)}
  (\bibinfo{year}{2015}).

\bibitem[{\citenamefont{{Mo} and {White}}(1996)}]{mw1996}
\bibinfo{author}{\bibfnamefont{H.~J.} \bibnamefont{{Mo}}} \bibnamefont{and}
  \bibinfo{author}{\bibfnamefont{S.~D.~M.} \bibnamefont{{White}}},
  \bibinfo{journal}{\mnras} \textbf{\bibinfo{volume}{282}},
  \bibinfo{pages}{347} (\bibinfo{year}{1996}), \eprint{astro-ph/9512127}.

\bibitem[{\citenamefont{{Sheth} and {Tormen}}(1999)}]{st1999}
\bibinfo{author}{\bibfnamefont{R.~K.} \bibnamefont{{Sheth}}} \bibnamefont{and}
  \bibinfo{author}{\bibfnamefont{G.}~\bibnamefont{{Tormen}}},
  \bibinfo{journal}{\mnras} \textbf{\bibinfo{volume}{308}},
  \bibinfo{pages}{119} (\bibinfo{year}{1999}), \eprint{astro-ph/9901122}.

\bibitem[{\citenamefont{Villaescusa-Navarro
  et~al.}(2018)\citenamefont{Villaescusa-Navarro, Banerjee, Dalal, Castorina,
  Scoccimarro, Angulo, and Spergel}}]{hades}
\bibinfo{author}{\bibfnamefont{F.}~\bibnamefont{Villaescusa-Navarro}},
  \bibinfo{author}{\bibfnamefont{A.}~\bibnamefont{Banerjee}},
  \bibinfo{author}{\bibfnamefont{N.}~\bibnamefont{Dalal}},
  \bibinfo{author}{\bibfnamefont{E.}~\bibnamefont{Castorina}},
  \bibinfo{author}{\bibfnamefont{R.}~\bibnamefont{Scoccimarro}},
  \bibinfo{author}{\bibfnamefont{R.}~\bibnamefont{Angulo}}, \bibnamefont{and}
  \bibinfo{author}{\bibfnamefont{D.~N.} \bibnamefont{Spergel}},
  \bibinfo{journal}{The Astrophysical Journal} \textbf{\bibinfo{volume}{861}},
  \bibinfo{pages}{53} (\bibinfo{year}{2018}),
  \urlprefix\url{https://doi.org/10.3847/1538-4357/aac6bf}.

\bibitem[{\citenamefont{{Cooray} and {Sheth}}(2002)}]{halomodel}
\bibinfo{author}{\bibfnamefont{A.}~\bibnamefont{{Cooray}}} \bibnamefont{and}
  \bibinfo{author}{\bibfnamefont{R.}~\bibnamefont{{Sheth}}},
  \bibinfo{journal}{Phys. Rep.} \textbf{\bibinfo{volume}{372}},
  \bibinfo{pages}{1} (\bibinfo{year}{2002}), \eprint{astro-ph/0206508}.

\bibitem[{\citenamefont{{Nusser} and {Davis}}(1994)}]{nd1994}
\bibinfo{author}{\bibfnamefont{A.}~\bibnamefont{{Nusser}}} \bibnamefont{and}
  \bibinfo{author}{\bibfnamefont{M.}~\bibnamefont{{Davis}}},
  \bibinfo{journal}{\apjl} \textbf{\bibinfo{volume}{421}}, \bibinfo{pages}{L1}
  (\bibinfo{year}{1994}), \eprint{astro-ph/9309009}.

\bibitem[{\citenamefont{{Fry}}(1996)}]{fry1996}
\bibinfo{author}{\bibfnamefont{J.~N.} \bibnamefont{{Fry}}},
  \bibinfo{journal}{\apjl} \textbf{\bibinfo{volume}{461}}, \bibinfo{pages}{L65}
  (\bibinfo{year}{1996}).

\bibitem[{\citenamefont{Bardeen et~al.}(1986)\citenamefont{Bardeen, Bond,
  Kaiser, and Szalay}}]{bbks86}
\bibinfo{author}{\bibfnamefont{J.~M.} \bibnamefont{Bardeen}},
  \bibinfo{author}{\bibfnamefont{J.~R.} \bibnamefont{Bond}},
  \bibinfo{author}{\bibfnamefont{N.}~\bibnamefont{Kaiser}}, \bibnamefont{and}
  \bibinfo{author}{\bibfnamefont{A.~S.} \bibnamefont{Szalay}},
  \bibinfo{journal}{Astrophys. J.} \textbf{\bibinfo{volume}{304}},
  \bibinfo{pages}{15} (\bibinfo{year}{1986}).

\bibitem[{\citenamefont{{Desjacques} et~al.}(2010)\citenamefont{{Desjacques},
  {Crocce}, {Scoccimarro}, and {Sheth}}}]{bkPeaks}
\bibinfo{author}{\bibfnamefont{V.}~\bibnamefont{{Desjacques}}},
  \bibinfo{author}{\bibfnamefont{M.}~\bibnamefont{{Crocce}}},
  \bibinfo{author}{\bibfnamefont{R.}~\bibnamefont{{Scoccimarro}}},
  \bibnamefont{and} \bibinfo{author}{\bibfnamefont{R.~K.}
  \bibnamefont{{Sheth}}}, \bibinfo{journal}{\prd}
  \textbf{\bibinfo{volume}{82}}, \bibinfo{eid}{103529} (\bibinfo{year}{2010}),
  \eprint{1009.3449}.

\bibitem[{\citenamefont{{Baldauf} and {Desjacques}}(2017)}]{baoPeaks}
\bibinfo{author}{\bibfnamefont{T.}~\bibnamefont{{Baldauf}}} \bibnamefont{and}
  \bibinfo{author}{\bibfnamefont{V.}~\bibnamefont{{Desjacques}}},
  \bibinfo{journal}{\prd} \textbf{\bibinfo{volume}{95}}, \bibinfo{eid}{043535}
  (\bibinfo{year}{2017}), \eprint{1612.04521}.

\bibitem[{\citenamefont{{Sarpa} et~al.}(2019)\citenamefont{{Sarpa}, {Schimd},
  {Branchini}, and {Matarrese}}}]{eFAM2019}
\bibinfo{author}{\bibfnamefont{E.}~\bibnamefont{{Sarpa}}},
  \bibinfo{author}{\bibfnamefont{C.}~\bibnamefont{{Schimd}}},
  \bibinfo{author}{\bibfnamefont{E.}~\bibnamefont{{Branchini}}},
  \bibnamefont{and}
  \bibinfo{author}{\bibfnamefont{S.}~\bibnamefont{{Matarrese}}},
  \bibinfo{journal}{\mnras} \textbf{\bibinfo{volume}{484}},
  \bibinfo{pages}{3818} (\bibinfo{year}{2019}), \eprint{1809.10738}.

\bibitem[{\citenamefont{{Anselmi} et~al.}(2016)\citenamefont{{Anselmi},
  {Starkman}, and {Sheth}}}]{rLP2016}
\bibinfo{author}{\bibfnamefont{S.}~\bibnamefont{{Anselmi}}},
  \bibinfo{author}{\bibfnamefont{G.~D.} \bibnamefont{{Starkman}}},
  \bibnamefont{and} \bibinfo{author}{\bibfnamefont{R.~K.}
  \bibnamefont{{Sheth}}}, \bibinfo{journal}{\mnras}
  \textbf{\bibinfo{volume}{455}}, \bibinfo{pages}{2474} (\bibinfo{year}{2016}),
  \eprint{1508.01170}.

\bibitem[{\citenamefont{{Anselmi} et~al.}(2018)\citenamefont{{Anselmi},
  {Corasaniti}, {Starkman}, {Sheth}, and {Zehavi}}}]{lpMocks}
\bibinfo{author}{\bibfnamefont{S.}~\bibnamefont{{Anselmi}}},
  \bibinfo{author}{\bibfnamefont{P.-S.} \bibnamefont{{Corasaniti}}},
  \bibinfo{author}{\bibfnamefont{G.~D.} \bibnamefont{{Starkman}}},
  \bibinfo{author}{\bibfnamefont{R.~K.} \bibnamefont{{Sheth}}},
  \bibnamefont{and} \bibinfo{author}{\bibfnamefont{I.}~\bibnamefont{{Zehavi}}},
  \bibinfo{journal}{\prd} \textbf{\bibinfo{volume}{98}}, \bibinfo{eid}{023527}
  (\bibinfo{year}{2018}), \eprint{1711.09063}.

\bibitem[{\citenamefont{{Nikakhtar} et~al.}(2021)\citenamefont{{Nikakhtar},
  {Sheth}, and {Zehavi}}}]{laguerre}
\bibinfo{author}{\bibfnamefont{F.}~\bibnamefont{{Nikakhtar}}},
  \bibinfo{author}{\bibfnamefont{R.~K.} \bibnamefont{{Sheth}}},
  \bibnamefont{and} \bibinfo{author}{\bibfnamefont{I.}~\bibnamefont{{Zehavi}}},
  \bibinfo{journal}{\prd} \textbf{\bibinfo{volume}{104}}, \bibinfo{eid}{043530}
  (\bibinfo{year}{2021}), \eprint{2101.08376}.

\bibitem[{\citenamefont{{Nikakhtar} et~al.}(2022)\citenamefont{{Nikakhtar},
  {Sheth}, and {Zehavi}}}]{hermite}
\bibinfo{author}{\bibfnamefont{F.}~\bibnamefont{{Nikakhtar}}},
  \bibinfo{author}{\bibfnamefont{R.~K.} \bibnamefont{{Sheth}}},
  \bibnamefont{and} \bibinfo{author}{\bibfnamefont{I.}~\bibnamefont{{Zehavi}}},
  \bibinfo{journal}{\prd} \bibinfo{eid}{043536} (\bibinfo{year}{2022}),
  \eprint{2110.03591}.

\bibitem[{\citenamefont{Bernardeau et~al.}(2002)\citenamefont{Bernardeau,
  Colombi, Gaztanaga, and Scoccimarro}}]{Bernardeau:2001qr}
\bibinfo{author}{\bibfnamefont{F.}~\bibnamefont{Bernardeau}},
  \bibinfo{author}{\bibfnamefont{S.}~\bibnamefont{Colombi}},
  \bibinfo{author}{\bibfnamefont{E.}~\bibnamefont{Gaztanaga}},
  \bibnamefont{and}
  \bibinfo{author}{\bibfnamefont{R.}~\bibnamefont{Scoccimarro}},
  \bibinfo{journal}{Phys. Rept.} \textbf{\bibinfo{volume}{367}},
  \bibinfo{pages}{1} (\bibinfo{year}{2002}), \eprint{astro-ph/0112551}.

\bibitem[{\citenamefont{{Despali} et~al.}(2013)\citenamefont{{Despali},
  {Tormen}, and {Sheth}}}]{shapes2013}
\bibinfo{author}{\bibfnamefont{G.}~\bibnamefont{{Despali}}},
  \bibinfo{author}{\bibfnamefont{G.}~\bibnamefont{{Tormen}}}, \bibnamefont{and}
  \bibinfo{author}{\bibfnamefont{R.~K.} \bibnamefont{{Sheth}}},
  \bibinfo{journal}{\mnras} \textbf{\bibinfo{volume}{431}},
  \bibinfo{pages}{1143} (\bibinfo{year}{2013}), \eprint{1212.4157}.

\bibitem[{\citenamefont{{Ludlow} et~al.}(2014)\citenamefont{{Ludlow},
  {Borzyszkowski}, and {Porciani}}}]{lbp2014}
\bibinfo{author}{\bibfnamefont{A.~D.} \bibnamefont{{Ludlow}}},
  \bibinfo{author}{\bibfnamefont{M.}~\bibnamefont{{Borzyszkowski}}},
  \bibnamefont{and}
  \bibinfo{author}{\bibfnamefont{C.}~\bibnamefont{{Porciani}}},
  \bibinfo{journal}{\mnras} \textbf{\bibinfo{volume}{445}},
  \bibinfo{pages}{4110} (\bibinfo{year}{2014}).

\bibitem[{\citenamefont{{Mohayaee} and {Tully}}(2005)}]{mt2005}
\bibinfo{author}{\bibfnamefont{R.}~\bibnamefont{{Mohayaee}}} \bibnamefont{and}
  \bibinfo{author}{\bibfnamefont{R.~B.} \bibnamefont{{Tully}}},
  \bibinfo{journal}{\apjl} \textbf{\bibinfo{volume}{635}},
  \bibinfo{pages}{L113} (\bibinfo{year}{2005}), \eprint{astro-ph/0509313}.

\bibitem[{\citenamefont{{Lavaux} et~al.}(2010)\citenamefont{{Lavaux}, {Tully},
  {Mohayaee}, and {Colombi}}}]{2massMAK}
\bibinfo{author}{\bibfnamefont{G.}~\bibnamefont{{Lavaux}}},
  \bibinfo{author}{\bibfnamefont{R.~B.} \bibnamefont{{Tully}}},
  \bibinfo{author}{\bibfnamefont{R.}~\bibnamefont{{Mohayaee}}},
  \bibnamefont{and}
  \bibinfo{author}{\bibfnamefont{S.}~\bibnamefont{{Colombi}}},
  \bibinfo{journal}{\apj} \textbf{\bibinfo{volume}{709}}, \bibinfo{pages}{483}
  (\bibinfo{year}{2010}), \eprint{0810.3658}.

\bibitem[{\citenamefont{{Cai} et~al.}(2011)\citenamefont{{Cai}, {Bernstein},
  and {Sheth}}}]{cai2011}
\bibinfo{author}{\bibfnamefont{Y.-C.} \bibnamefont{{Cai}}},
  \bibinfo{author}{\bibfnamefont{G.}~\bibnamefont{{Bernstein}}},
  \bibnamefont{and} \bibinfo{author}{\bibfnamefont{R.~K.}
  \bibnamefont{{Sheth}}}, \bibinfo{journal}{\mnras}
  \textbf{\bibinfo{volume}{412}}, \bibinfo{pages}{995} (\bibinfo{year}{2011}),
  \eprint{1007.3500}.

\bibitem[{\citenamefont{Wang et~al.}(2014)\citenamefont{Wang, Mo, Yang, Jing,
  and Lin}}]{elucid2014}
\bibinfo{author}{\bibfnamefont{H.}~\bibnamefont{Wang}},
  \bibinfo{author}{\bibfnamefont{H.~J.} \bibnamefont{Mo}},
  \bibinfo{author}{\bibfnamefont{X.}~\bibnamefont{Yang}},
  \bibinfo{author}{\bibfnamefont{Y.~P.} \bibnamefont{Jing}}, \bibnamefont{and}
  \bibinfo{author}{\bibfnamefont{W.~P.} \bibnamefont{Lin}},
  \bibinfo{journal}{The Astrophysical Journal} \textbf{\bibinfo{volume}{794}},
  \bibinfo{pages}{94} (\bibinfo{year}{2014}),
  \urlprefix\url{https://doi.org/10.1088/0004-637x/794/1/94}.

\bibitem[{\citenamefont{{Paranjape} and {Alam}}(2020)}]{voronoi2020}
\bibinfo{author}{\bibfnamefont{A.}~\bibnamefont{{Paranjape}}} \bibnamefont{and}
  \bibinfo{author}{\bibfnamefont{S.}~\bibnamefont{{Alam}}},
  \bibinfo{journal}{\mnras} \textbf{\bibinfo{volume}{495}},
  \bibinfo{pages}{3233} (\bibinfo{year}{2020}), \eprint{2001.08760}.

\end{thebibliography}

\end{document}